\documentclass[a4paper]{article}
\usepackage[latin1]{inputenc}
\usepackage[T1]{fontenc}
\usepackage[english]{babel}
\usepackage{amsmath}
\usepackage{amssymb,amsfonts,textcomp}
\usepackage{color}
\usepackage{array}
\usepackage{supertabular}
\usepackage{hhline}
\usepackage{wrapfig}
\usepackage{hyperref}
\usepackage{indentfirst}
\hypersetup{pdftex, colorlinks=true, linkcolor=blue, citecolor=blue, filecolor=blue, urlcolor=blue, pdftitle=, pdfauthor=, pdfsubject=, pdfkeywords=}
\usepackage{graphicx}
\newcommand\textsubscript[1]{\ensuremath{{}_{\text{#1}}}}
\makeatletter
\newcommand\arraybslash{\let\\\@arraycr}
\makeatother
\makeatletter
\newcommand\ps@Standard{
  \renewcommand\@oddhead{}
  \renewcommand\@evenhead{}
  \renewcommand\@oddfoot{}
  \renewcommand\@evenfoot{}
  \renewcommand\thepage{\arabic{page}}
}
\makeatother
\title{}
\author{}
\date{2013-07-12}
\begin{document}
\clearpage\setcounter{page}{1}\pagestyle{Standard}
\begin{flushleft}
{\selectlanguage{english}
\textbf{SInC: An accurate and fast error-model based simulator for SNPs, Indels
and CNVs coupled with a read generator for short-read sequence data }}
\end{flushleft}
\begin{flushleft}
{\selectlanguage{english}\bfseries
Authors}
\end{flushleft}

{\selectlanguage{english}
Swetansu Pattnaik\textsuperscript{1,2}, Saurabh Gupta\textsuperscript{1}, Arjun
A Rao\textsuperscript{1} and Binay Panda\textsuperscript{1,2*}}

\begin{flushleft}
{\selectlanguage{english}\bfseries
Affiliations}
\end{flushleft}

{\selectlanguage{english}
\textsuperscript{1} Ganit Labs, Bio-IT Centre, Institute of Bioinformatics and
Applied Biotechnology, Biotech Park, Electronic City Phase I, Bangalore 560100,
India}

{\selectlanguage{english}
\textsuperscript{2 }Strand Life Sciences, Kirloskar Business Park, Bellary Road,
Hebbal, Bangalore 560024, India}

{\selectlanguage{english}
* corresponding author (binay@ganitlabs.in)}

\begin{flushleft}
{\selectlanguage{english}\bfseries
Abstract}
\end{flushleft}

{\selectlanguage{english}
We report SInC (SNV, Indel and CNV) simulator and read generator, an open-source
tool capable of simulating biological variants taking into account a
platform-specific error model. SInC is capable of simulating and generating
single- and paired-end reads with user-defined insert size with high efficiency
compared to the other existing tools. SInC, due to its multi-threaded
capability during read generation, has a low time footprint. SInC is currently
optimised to work in limited infrastructure setup and can efficiently exploit
the commonly used quad-core desktop architecture to simulate short sequence
reads with deep coverage for large genomes.}
SInC can be downloaded from http://sourceforge.net/projects/sincsimulator.
\begin{flushleft}
{\selectlanguage{english}\bfseries
Introduction}
\end{flushleft}

{\selectlanguage{english}
The rapid advancements in the field of genome sequencing is aiding our
understanding of genome organisation in many biological systems[1-3]. These
tools are intended to analyse high throughput next-generation sequence (NGS)
data and present biologically relevant interpretations. Given the high
throughput nature of present day genomics, heuristic algorithms are implicated
to identify or predict genome variations as small as single base nucleotide
substitutions (SNVs) to insertion-deletion events (indels) and copy number
variations (CNVs). Hence, it is imperative for developers of NGS data analysis
pipelines to establish the limits of their predictions based on simulated data
as in current practice. \foreignlanguage{english}{In the last five years,
computational biologists and bioinformatics specialists have developed new
algorithms for different types of variant calling, have implemented existing
algorithms for short-read mapping to reference genomes and/or optimized
pipelines to perform a specific type of primary and secondary analysis[4-19].
SNVs, indels and CNVs are the most common types of biological variations in the
genome. The tools to detect these variants have the common objective of finding
novel variations with low frequency of false positives,
}\foreignlanguage{english}{rediscovering known variations in the genome of
interest and facilitate subsequent genome visualization and interpretation.
Hence, availability of reliable and realistic simulated dataset bearing the
three major types of genomic variations (SNVs, indels and CNVs) is critical to
test the operational limitations of newly developed or existing tools. This
approach allows computational biologists to generate simulated datasets with
biological meaning and sensitive to systematic error inherent to different
sequencing technology platforms.}}

{\selectlanguage{english}
\foreignlanguage{english}{Although, next-generation sequencing (NGS) instruments
generate reads of various lengths and with varying error profiles, the most
popular source of data remains sequencing instruments from Illumina, which
employs a sequencing-by-synthesis chemistry to generate short-reads. Keeping
this in mind, we have developed an efficient, fast simulator and read generator
that mimics sequencing errors generated by Illumina platform. The tool was
developed using the Illumina error model to cater to the larger interest group.
However, it can easily be adopted for any other sequencing platform by
supplying an instrument-specific error profile.}}

{\selectlanguage{english}
\foreignlanguage{english}{Currently available tools can either generate
platform-specific, error-profile based reads or simulate reads across
platforms[21-28]. It is also in our interest of disambiguation to classify the
existing simulators into two major classes based on their functionality. First,
the stand-alone read generators (RG) like Metasim[28], Flowsim[22], 454Sim[24],
Pbsim[21], GenFrag[29] and ART[25] among others with functionality limited to
read generation. The second class of simulators (SRG) include pIRS[23],
GEMsim[26], dwgSIM[30] (based on wgsim of samtools) which have the option of
simulating genomic variations coupled with read generation functionality. Each
of the above mentioned tools, although has its own set of advantages, suffers
from either having a simplistic error model (in the case of GenFrag), errors
that does not model real data (in the case of dwgSIM), does not assign quality
values to reads (in the case of Metasim), does not simulate Illumina reads (in
the case of Flowsim) or does not simulate multiple types of variations (in the
case of pIRS and GEMsim). Interestingly, none of the existing SRG simulators
present the option to simulate CNVs. Hence, we have developed and implemented a
C-program, SInC, to enable simulation of all the three major types of genomic
variations, SNVs, indels and CNVs, coupled with a multi-threaded, error-profile
based read generator. SInC has obvious advantages over the popular SRG
simulators as dwgSIM simulates reads with identical dummy base quality values
relieving the data of any base-quality related effects, pIRS cannot simulate
CNVs and GEMsim simulates only SNVs. SInC models errors based on real data from
Illumina instruments as in pIRS and additionally presents fine tuned options to
replicate biologically meaningful variant simulations including CNVs. The
multi-threaded algorithm in SInC for read generation provides substantial
advantage in run time and allows for seamless simulation of high coverage data
in a desktop environment.}}

{\selectlanguage{english}
\foreignlanguage{english}{Here we present an evaluation of SInC using commonly
used SNV, indel and CNV detection tools. The speed, accuracy and efficiency was
compared against other popular simulators and read generators.}}
\begin{flushleft}
{\selectlanguage{english}\bfseries
Methods}
\end{flushleft}

{\selectlanguage{english}
SInC performs two jobs; first it simulates variants (simulator) and then it
generates reads (read generator). SInC simulator consists of three independent
modules (one each for SNV, indel and CNV) that can either be executed
independently in a mutually exclusive manner or in any combination.}
\begin{flushleft}
{\selectlanguage{english}\bfseries
SNV simulation}
\end{flushleft}

{\selectlanguage{english}
The exact frequency of SNPs in the human genome has not yet been determined
accurately. Based on inferences from 629 complete genomes representing several
human populations in the 1000 genome data, the current range of frequency of
SNV lies between one per 300 to 1000 bases[31]. For this purpose, we have
assumed that the substitution events in human genome are independent and
random. SInC simulator accepts a user defined percentage value to simulate
SNVs. The algorithm identifies this percentage value as the fraction of genome
to estimate the number of SNVs and simulates SNVs across the genome. To
maintain positional identities of these SNVs with respect to their frequency,
that are normally distributed over the sequenced genome, the mean distance of
separation (\textit{D}\textit{\textsubscript{Avg}}) between SNVs
is\textsubscript{ }calculated (see \foreignlanguage{english}{Additional file
1and Additional file 2)}.}

{\selectlanguage{english}
This ensures that the simulated SNVs are well distributed over the genome. A
positional filter is applied to remove the outlier SNVs, which are less than 15
bases apart. SInC simulator neglects SNVs simulated in the N-regions of the
genome (where there is no A, T, G or C). Then the algorithm applies a
user-defined transition to transversion (Ti/Tv) metric to maintain the
biological significance of the SNVs across the genome. A Ti/Tv ratio of 2.1 was
maintained across the population of simulated SNVs with 20\% inherent
heterozygosity to simulate human genome data as previously reported[32]. The
flow chart illustrated in Figure 1A depicts the algorithm for simulating SNVs.}
\begin{flushleft}
{\selectlanguage{english}\bfseries
Indel simulation}
\end{flushleft}

{\selectlanguage{english}
Insertion and deletion (indel) events have a wide range of size-based
variability. SInC simulator simulates short, medium and large indels in the
range of 1-10bp 11-20bp and 21-100bp respectively in concordance with earlier
studies[33]. The ratio of incidence of insertions to deletions and heterozygous
to homozygous indels in human genome is set to 1:1 based on previous
observations[33]. The flow chart in Figure 1B depicts the algorithm for
simulating indels. }

{\selectlanguage{english}
\foreignlanguage{english}{The algorithm first randomly generates the position
for indels and then uses a filter to replace any indel within the region of the
N-region of the genome (no A, T, G or C assigned) with one in the sequenced
region. To remove duplicates, the simulated indels are coordinate sorted and
only the unique locations are retained. Usually, a redundancy of 2-5\% is
filtered out post }\foreignlanguage{english}{coordinate sorting (that can
result from either duplicates or un-sequenced regions). Hence, an additional
5\% of indels are generated at the initial stage of the algorithm to account
for the loss of indels at the duplicate removal step. In the next stage of the
algorithm, the frequency of short, medium and large indels was factored in
based on previous literature evidence for their distribution in human[33]. }The
indel simulation produces two output files assuming the bi-allelic nature of
human genes, each containing allele-specific coordinate information of
simulated indels. \foreignlanguage{english}{Among the total number of simulated
indels, the algorithm simulates 30\% single base indels, 20\% repeat
expansions, 49\% 2-20bp indels, and 1\% long indels including repeat expansions
(see Additional file 2 and Additional file 3).}}
\begin{figure}[h!]
\begin{center}
\includegraphics[scale=0.8]{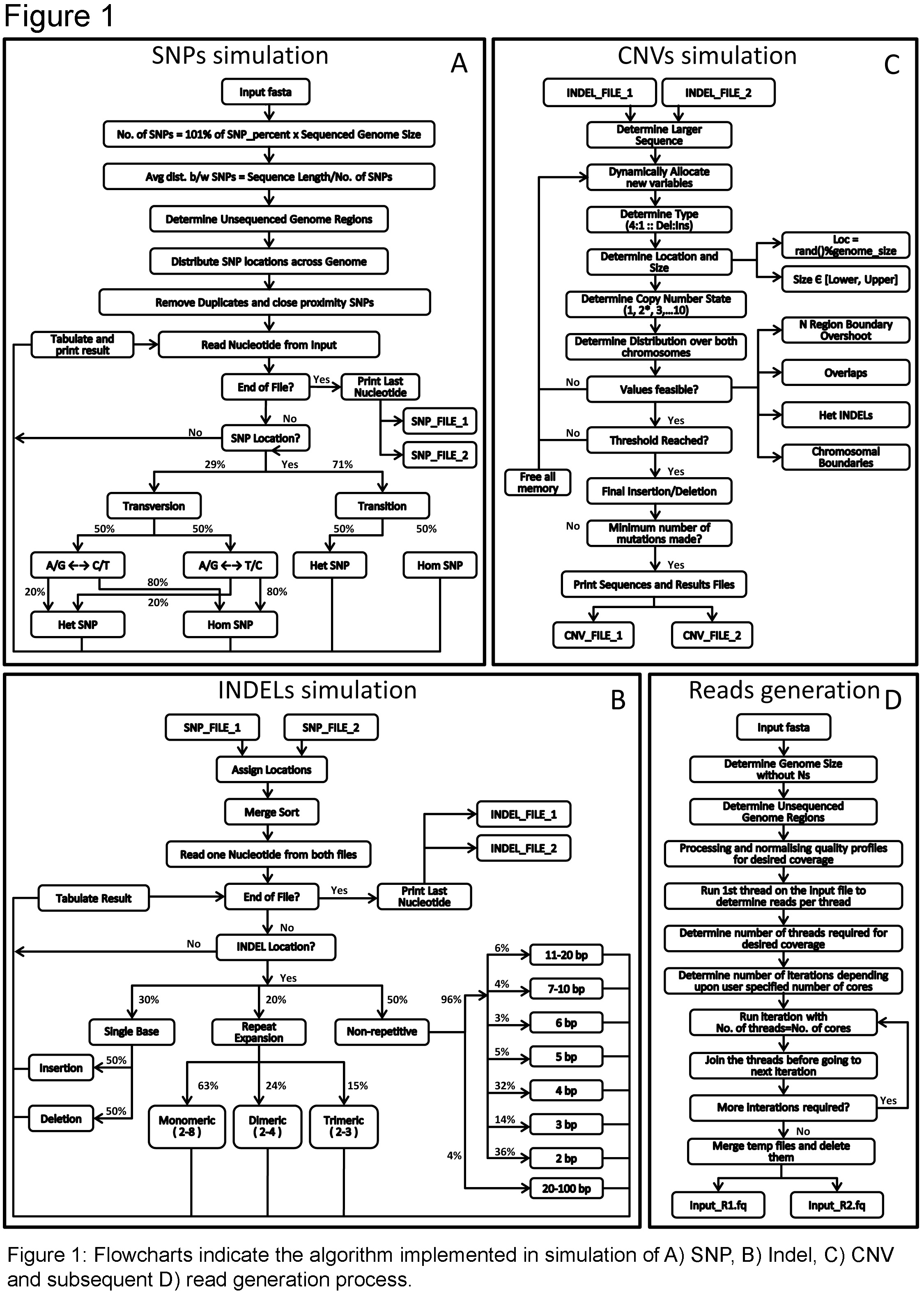}
\end{center}
\end{figure}
\begin{flushleft}
{\selectlanguage{english}\bfseries
CNV simulation}
\end{flushleft}

{\selectlanguage{english}
\foreignlanguage{english}{The CNV simulation constitutes the final step of the
simulation algorithm, as it can ply in a sequential manner post indel
simulation. Since the input files from indel simulation may contain
heterozygous indels, which may be of unequal size, hence the CNV module takes
it into consideration and prevents the possibility of boundary overlaps with
indels. }The flow chart in Figure 1C depicts the algorithm for simulating
CNVs.}

{\selectlanguage{english}
\foreignlanguage{english}{Unlike the indel module, here the size and location of
the variants are both generated dynamically with the flow of the program after
obtaining the feed from the user to determine the number of CNVs and their
range of size distribution (upper and lower limit). Such simulated data is
particularly useful to test the accuracy and sensitivity of a new or existing
CNV caller across a wide range of CNV sizes. The algorithm then filters the
simulated CNVs based on its coordinates. First the span of each of the CNVs are
evaluated to ensure correspondence with chromosomal boundary in either allele
and subsequently the CNV boundaries are checked for overlap with neighboring
CNVs. The CNV is logged and the next iteration of location and size are
generated upon meeting the aforementioned conditions. Unlike the SNV and indel
simulation modules, the annotation data for both alleles in the CNV module is
stored in the same file in the form of a tabular data. The tool also outputs a
simplified results file (similar to a BED file), which can be read easily by
any program for visualizing CNVs.}}
\begin{flushleft}
{\selectlanguage{english}\bfseries
Read Generation}
\end{flushleft}

{\selectlanguage{english}
\foreignlanguage{english}{SInC has a read generator part that generates short
reads using a multi-threaded approach. This uses a profile-based error-model
from Illumina instruments with paired-end reads and utilizes the parallel
processing power of commonly used quad-core desktop/laptop architecture. The
algorithm follows a {\textquotedblleft}divide and conquer{\textquotedblright}
approach where each thread spans the input sequence once and the number of
reads required to obtain the user defined coverage are pooled from the
estimated number of threads. User-specified cores utilization is implemented in
the SInC read generator to prevent over-utilization of available CPU resources.
The other major user defined parameters, include read length, error profile,
insert size (inner distance) and standard deviation of insert size [see
Additional file 2]. The algorithm initially creates one thread, which generates
reads for the input fasta file. }\foreignlanguage{english}{Depending on the
read pairs generated in the first run, the numbers of threads required to
obtain the desired coverage are calculated and then executed in an iterative
manner based on the number of cores specified by the user (Figure 1D). SInC is
optimized to run with 4 threads suiting a quad-core processor. }}
\begin{flushleft}
{\selectlanguage{english}\bfseries
Evaluation of SInC:}
\end{flushleft}

\begin{flushleft}
{\selectlanguage{english}\bfseries
Variant re-discovery }
\end{flushleft}

{\selectlanguage{english}
\foreignlanguage{english}{We used human chromosome 22 sequence from the UCSC
build hg19 for generating SNVs and indels using all the four different SRG
simulators. The SNV rate, indel percentage and coverage was maintained across
all the tools and the resulting reads were aligned using Novoalign[14]. These
mapped files were subject to SNV and indel detection by GATK[4] and Pindel[6]
respectively (see Results, Figure 2). The predicted SNVs and indels from the different
simulators were compared to the actual number of incorporated variants to
estimate the percentage rediscovery. Rediscovery percentage using Pindel has a
limitation that it merges short indels within a span of 40 nucleotides of each
other leading to a slight loss (less than 1\%) of rediscovered indels across
all the simulators (see Additional file 2).}}
\begin{flushleft}
{\selectlanguage{english}\bfseries
Time profiling}
\end{flushleft}

{\selectlanguage{english}
\foreignlanguage{english}{Given the high-throughput nature of NGS data,
generating the bulk of simulated data still remains a time consuming process.
Hence, we have implemented a {\textquotedblleft}divide and
conquer{\textquotedblright} approach to the read generation module to reduce
the time footprint in generating high coverage data. This property allows user
to simulate data at a high coverage (50X -- 100X) without inordinate expense of
time. SInC can utilize 1 to 4 threads for optimal function. Our comparison was
set up based on default use of 1 core ranging upto a maximum utilization of 4
cores in SInC versus the other tools (see Results, Figure 3). Details are provided in the
Additional file 2. }}
\begin{flushleft}
{\selectlanguage{english}
\textbf{Transition/Transversion (Ti/Tv) ratio}}
\end{flushleft}

{\selectlanguage{english}
A transition mutation involves a change from purine to purine or pyrimidine to
pyrimidine and a tranversion mutation involves a change from pyrimidine to
purine or \textit{vice versa}. This makes a transversion event twice as
favourable as a transition event for any random mutation event. Hence, the
Ti/Tv ratio for a random variation resulting from systematic errors in the
sequencing technology, alignment artefacts and data processing failures should
be close to 0.5. As published earlier, Ti/Tv ratio for whole genome falls
between 2.05 - 2.15 for both known and unknown SNPs. SInC incorporates a
user-defined Ti/Tv ratio for simulation of SNVs.}

{\selectlanguage{english}
The scripts to simulate variants and generate reads used here are given in the
Additional file 4.}
\begin{flushleft}
{\selectlanguage{english}
\foreignlanguage{english}{\textbf{Results}}\foreignlanguage{english}{ }}
\end{flushleft}
\begin{figure}[b!]
\centering
\includegraphics[scale=0.45]{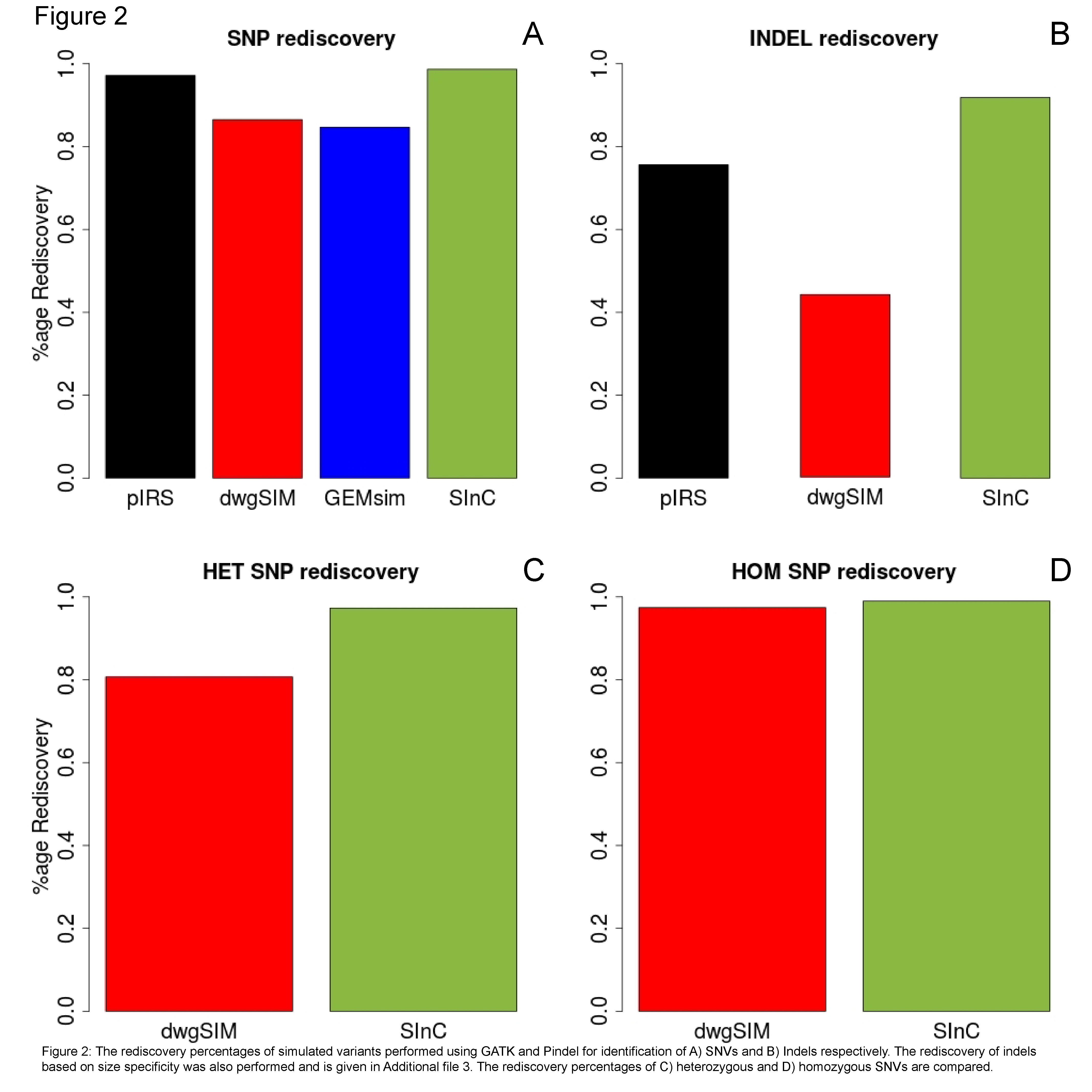}
\end{figure}
\begin{figure}[b!]
\centering
\includegraphics[scale=0.45]{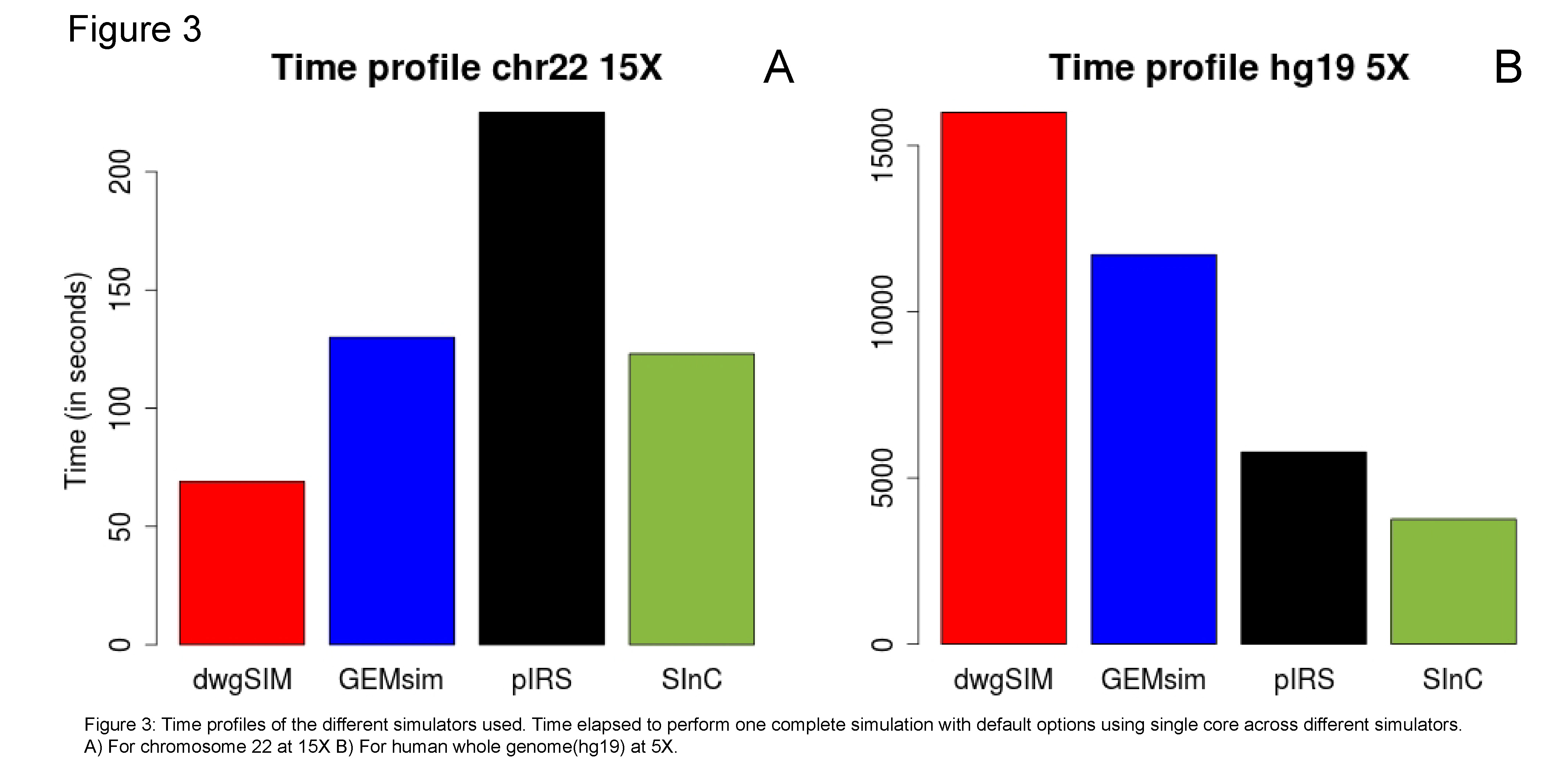}
\end{figure}
{\selectlanguage{english}
\foreignlanguage{english}{We have developed a simulator for all commonly
occurring biological variants in the genome }\foreignlanguage{english}{along
with a read generator. We compared the latest pick of simulators with SInC
simulator and read generator. In our model for SNV simulation, we have limited
the range of simulation of SNVs using a distribution of distance between two
consecutive SNVs. Based on SNV frequency studies in human genome[31, 34], under
default simulation parameters the mean distance between two consecutive SNVs,
}\foreignlanguage{english}{\textit{D}}\foreignlanguage{english}{\textit{\textsubscript{Avg}}}\foreignlanguage{english}{,
is set dynamically between 300 to 1000 bases depending upon user defined input
for SNV rate. In indel simulations, the complexity of simulation depends on the
frequency of indels in the simulated data. In the default mode for indel
simulation, the algorithm is sensitive to the natural frequencies and size
ranges as evidenced from existing literature[33, 35]. The model for CNV
simulation is an extension of the indel simulation, wherein the CNVs are
dynamically generated while maintaining the allele specificity and genomic
positions of indels simulated in the prior step. The simulated variants are
captured in log files, combined with input allele fasta files and processed by
a multi-threaded process to enable fast-paced read generation.}}

{\selectlanguage{english}
\foreignlanguage{english}{In order to assess the efficiency of SInC in
comparison to existing simulators, we compared variant re-discovery rate and
time taken to complete the job among all the tools. We implemented a variant
re-discovery strategy employing GATK[4] for SNV detection and Pindel[6] for
detecting indels. The SNV rediscovery percentage suggested that SInC was at par
with pIRS in the efficiency of simulating SNVs and comprehensively outperformed
both GEMsim and dwgSIM (Figure 2A). Although other tools like dwgSIM and GEMsim
are close to SInC in homozygous rediscovery rate (Figure 2D), SInC outperforms
both these tools for heterozygous rediscovery rate (Figure 2C) suggesting the
importance in simulating both homozygous and heterozygous real variants. In
the rediscovery of indels, SInC emerges as the only simulator with the highest
percentage of total rediscovered indels, ahead at least by 15 \% from the
closest contender pIRS (Figure 2B). We further tested the accuracy of the
rediscovered indels by adding a size-based constraint and estimated the
percentage rediscovered in the size ranges containing 1 to 6, 7 to 10, 11 to 20
and 21 to 100 nucleotide long indels. These size ranges were simulated due to
their overall high (greater than 95\%) natural prevalence in human genomes.
This exercise corroborated the superiority of SInC in detecting indels while
retaining the size-based constraints implicated in the simulation algorithm in
comparison to the other tools tested (Figure 2B). The numbers of SNVs and
indels rediscovered by SInC are especially important because the total number
of SNVs simulated by SInC is about 10-20\% more than the other tools tested and
20-40\% higher for indels. Another significant advantage of SInC is apparent
from the rediscovered heterozygous SNVs. As depicted in Figure 2C, the
difference in homozygous SNV rediscovery is rather conserved across the
simulators compared to Figure 2D, which gives SInC an edge in conservation of
zygosity of the calls post read generation. Notably, pIRS although has a SNV
rediscovered percentage at par with SInC, it does not catalog the simulated
SNVs to facilitate rediscovery of heterozygous and homozygous SNVs separately.
The CNV module of SInC simulator was used in a previous study to test a CNV
prediction tool, COPS[17], and was used to compare its accuracy and sensitivity
to other popular CNV prediction tools. We were unable to perform a comparative
analysis of the CNV module in SInC due to the
}\foreignlanguage{english}{unavailability of any published tools that can
simulate CNVs. However, as previously shown[17], the percentage rediscovery
using multiple CNV discovery tools like CNAseg CNV-seq, CNVnator and SVDetect
yielded {\textgreater}90\% CNVs. }}

{\selectlanguage{english}
\foreignlanguage{english}{Next, we wanted to test the speed of SInC read
generator. Figure 3 depicts the advantages that SInC provides during read
generation due to implementation of a {\textquotedblleft}divide and
conquer{\textquotedblright} approach by efficient utilization of C thread
functions. The tool was tested for its processing capability under a range of
multi-threaded options ranging from default utilization of 1 core to a maximum
utilization of 4 cores. SInC accomplished read generation at least one and a
half times faster than pIRS and three times faster than ART; the two most
recent Illumina read simulators [see Supplementary Tables]. The time profile
demonstrated substantial reduction in time footprint using SInC in comparison
to the other tools sampled in our study. This difference in generation time of
simulated data is reflected clearly in generating high coverage datasets from
large genomes, human genome in our case as shown in Figures 3B and 3C.}}
\begin{flushleft}
{\selectlanguage{english}\bfseries
Discussion}
\end{flushleft}

{\selectlanguage{english}
Although there are a multitude of popular tools capable of predicting genomic
variations using high-throughput sequence data, the generality of such tools
are questionable. In many ways, a simulated dataset is crucial towards
determining the success of predictive algorithms in the context of real
dataset. Simulators that can simulate variants and
\foreignlanguage{english}{generate reads are valuable tools used for developing
and testing tools for sequence data analysis. An ideal tool that can both
simulate multiple variant types (SNVs, indels and CNVs) and generate sequencing
reads taking into account a realistic platform-specific error-profile of an
sequencing instrument is currently lacking. We tried filling this void by
designing a versatile and fast tool that can generate multiple types of
biological variants (SNVs, indels and CNVs) and can run on a minimalistic quad
core desktop computer using multi-threaded option. The time advantage obtained
in SInC could be attributed to the optimized algorithms and efficient use of C
thread functions to manage the I/O streams. This advantage is also obvious in a
single core, which delegates the bulk of the data generation to multiple
threads to ensure efficient use of memory in line with
{\textquotedblleft}divide and conquer{\textquotedblright} approach. The
optimization of multiple core usage is available upto 4 cores in quad-core
architecture.}}

{\selectlanguage{english}
\foreignlanguage{english}{Another major functional advantage of this tool is its
ability to simulate CNVs. CNVs have been shown to contribute more towards
genetic diversity than SNVs and are conspicuous by their pervasiveness in human
genome[36-39]. The advent of NGS platforms has geared multiple efforts to build
frameworks towards identifying CNVs and assess their penetrance in disease
etiology. However, most of these efforts are only partially effective in
capturing population-based generalizations. In order to build a robust and
generic framework, it is imperative to build exhaustive datasets with the known
signatures and explore the range of false discovery rates inherent to the tools
and subsequently improve them. The ability to create such datasets will
definitely improve the approach and accuracy of predictions made
}\foreignlanguage{english}{by existing tools. Hence, a flexible, user-input
based simulator has substantial application in building useful datasets
allowing for improvement of current approaches towards variant discovery as a
whole.}\foreignlanguage{english}{ }\foreignlanguage{english}{Although there
have been efforts in the past to discovering CNVs using NGS data, currently
there are no available simulators to fine-tune CNV detection algorithms. SInC
simulator not only fulfills the simulation of CNVs but an additional
functionality of SInC simulator is to generate allele-specific CNVs. This is
particularly useful if one has to understand the copy number changes at an
allelic level important for many diseases[40, 41]. }}

{\selectlanguage{english}
\foreignlanguage{english}{Production of large amount of heterogeneous data in
high-throughput biology requires sophisticated computational tools for
efficient analysis, storage, sharing and archiving. This requires resources,
both software and hardware, and interoperability of computational
resources.}\foreignlanguage{english}{ A common practice among computational
biologist is to use simulated data to test the efficacy of the tools before
applying them to real dataset. Although there are many simulators available
currently, there is none that suits the need of every computational biologist
wanting to make tools for short-read sequence data. Keeping this in mind, we
have developed a tool to help computational biologists create simulated
datasets using only one simulator that can span across sequencing platforms and
variant types (SNVs, indels and CNVs). }\foreignlanguage{english}{Although,
SInC simulator was tested with human genome, it is versatile to address the
complexity of any genome, its substitution rate, variant frequency and
transition to transversion ratio. Large genomes, like that from many plants,
need time to generate simulated reads at high coverage and this is where the
multi-threaded capability of SInC scores high in comparison to other tools. }}

{\selectlanguage{english}
\foreignlanguage{english}{In conclusion, the ability of SInC to generate
realistic fastq reads based on Illumina read error profiles along with the
capacity to simulate multiple biological variants and generate reads
concurrently makes it a powerful option in a variety of simulation studies.}}
\begin{flushleft}
{\selectlanguage{english}\bfseries
Acknowledgement}
\end{flushleft}

{\selectlanguage{english}
We thank Professor N.Yathindra for encouragement.}
\begin{flushleft}
{\selectlanguage{english}\bfseries
Funding}
\end{flushleft}

{\selectlanguage{english}
Research is funded by Department of Electronics and Information Technology,
Government of India (Ref No:18(4)/2010-E-Infra., 31-03-2010) and Department of
IT, BT and ST, Government of Karnataka, India (Ref No:3451-00-090-2-22) \ under
the {\textquotedblleft}Bio-IT Project{\textquotedblright}.}
\begin{flushleft}
{\selectlanguage{english}\bfseries
References}
\end{flushleft}

{\selectlanguage{english}
1.~~~~~~~~ Schweiger MR, Kerick M, Timmermann B, Isau M: \textbf{The power of
NGS technologies to delineate the genome organization in cancer: from mutations
to structural variations and epigenetic alterations}. \textit{Cancer metastasis
reviews }2011, \textbf{30}(2):199-210.}

{\selectlanguage{english}
2.~~~~~~~~ Shendure J, Ji H: \textbf{Next-generation DNA sequencing}.
\textit{Nature biotechnology }2008, \textbf{26}(10):1135-1145.}

{\selectlanguage{english}
3.~~~~~~~~ Shendure J, Lieberman Aiden E: \textbf{The expanding scope of DNA
sequencing}. \textit{Nature biotechnology }2012, \textbf{30}(11):1084-1094.}

{\selectlanguage{english}
4.~~~~~~~~ McKenna A, Hanna M, Banks E, Sivachenko A, Cibulskis K, Kernytsky A,
Garimella K, Altshuler D, Gabriel S, Daly M \textit{et al}: \textbf{The Genome
Analysis Toolkit: a MapReduce framework for analyzing next-generation DNA
sequencing data}. \textit{Genome research }2010, \textbf{20}(9):1297-1303.}

{\selectlanguage{english}
5.~~~~~~~~ Li H: \textbf{A statistical framework for SNP calling, mutation
discovery, association mapping and population genetical parameter estimation
from sequencing data}. \textit{Bioinformatics }2011,
\textbf{27}(21):2987-2993.}

{\selectlanguage{english}
6.~~~~~~~~ Ye K, Schulz MH, Long Q, Apweiler R, Ning Z: \textbf{Pindel: a
pattern growth approach to detect break points of large deletions and medium
sized insertions from paired-end short reads}. \textit{Bioinformatics }2009,
\textbf{25}(21):2865-2871.}

{\selectlanguage{english}
7.~~~~~~~~ Albers CA, Lunter G, MacArthur DG, McVean G, Ouwehand WH, Durbin R:
\textbf{Dindel: accurate indel calls from short-read data}. \textit{Genome
research }2011, \textbf{21}(6):961-973.}

{\selectlanguage{english}
8.~~~~~~~~ Pattnaik S, Vaidyanathan S, Pooja DG, Deepak S, Panda B:
\textbf{Customisation of the exome data analysis pipeline using a combinatorial
approach}. \textit{PloS one }2012, \textbf{7}(1):e30080.}

{\selectlanguage{english}
9.~~~~~~~~ Li H, Durbin R: \textbf{Fast and accurate short read alignment with
Burrows-Wheeler transform}. \textit{Bioinformatics }2009,
\textbf{25}(14):1754-1760.}

{\selectlanguage{english}
10.~~~~~~ Homer N, Nelson SF: \textbf{Improved variant discovery through local
re-alignment of short-read next-generation sequencing data using SRMA}.
\textit{Genome biology }2010, \textbf{11}(10):R99.}

{\selectlanguage{english}
11.~~~~~~ Lunter G, Goodson M: \textbf{Stampy: a statistical algorithm for
sensitive and fast mapping of Illumina sequence reads}. \textit{Genome research
}2011, \textbf{21}(6):936-939.}

{\selectlanguage{english}
12.~~~~~~ Langmead B: \textbf{Aligning short sequencing reads with Bowtie}.
\textit{Current protocols in bioinformatics / editoral board, Andreas D
Baxevanis~ [et al] }2010, \textbf{Chapter 11}:Unit 11 17.}

{\selectlanguage{english}
13.~~~~~~ Liu Y, Schmidt B: \textbf{Long read alignment based on maximal exact
match seeds}. \textit{Bioinformatics }2012, \textbf{28}(18):i318-i324.}

{\selectlanguage{english}
\foreignlanguage{english}{14.~~~~~~ }\foreignlanguage{english}{\textbf{Novoalign
}}\textbf{Available: }\url{http://www.novocraft.com/main/index.php}\textbf{.
Accessed 2012 Dec.}}

{\selectlanguage{english}
15.~~~~~~ Ruffalo M, LaFramboise T, Koyuturk M: \textbf{Comparative analysis of
algorithms for next-generation sequencing read alignment}.
\textit{Bioinformatics }2011, \textbf{27}(20):2790-2796.}

{\selectlanguage{english}
16.~~~~~~ Hatem A, Bozdag D, Toland AE, Catalyurek UV: \textbf{Benchmarking
short sequence mapping tools}. \textit{BMC bioinformatics }2013,
\textbf{14}:184.}

{\selectlanguage{english}
17.~~~~~~ Krishnan NM, Gaur P, Chaudhary R, Rao AA, Panda B: \textbf{COPS: a
sensitive and accurate tool for detecting somatic Copy Number Alterations using
short-read sequence data from paired samples}. \textit{PloS one }2012,
\textbf{7}(10):e47812.}

{\selectlanguage{english}
18.~~~~~~ Abyzov A, Urban AE, Snyder M, Gerstein M: \textbf{CNVnator: an
approach to discover, genotype, and characterize typical and atypical CNVs from
family and population genome sequencing}. \textit{Genome research }2011,
\textbf{21}(6):974-984.}

{\selectlanguage{english}
19.~~~~~~ Xie C, Tammi MT: \textbf{CNV-seq, a new method to detect copy number
variation using high-throughput sequencing}. \textit{BMC bioinformatics }2009,
\textbf{10}:80.}

{\selectlanguage{english}
20.~~~~~~ Bentley DR, Balasubramanian S, Swerdlow HP, Smith GP, Milton J, Brown
CG, Hall KP, Evers DJ, Barnes CL, Bignell HR \textit{et al}: \textbf{Accurate
whole human genome sequencing using reversible terminator chemistry}.
\textit{Nature }2008, \textbf{456}(7218):53-59.}

{\selectlanguage{english}
21.~~~~~~ Ono Y, Asai K, Hamada M: \textbf{PBSIM: PacBio reads
simulator-{}-toward accurate genome assembly}. \textit{Bioinformatics }2013,
\textbf{29}(1):119-121.}

{\selectlanguage{english}
22.~~~~~~ Balzer S, Malde K, Lanzen A, Sharma A, Jonassen I:
\textbf{Characteristics of 454 pyrosequencing data-{}-enabling realistic
simulation with flowsim}. \textit{Bioinformatics }2010,
\textbf{26}(18):i420-425.}

{\selectlanguage{english}
23.~~~~~~ Hu X, Yuan J, Shi Y, Lu J, Liu B, Li Z, Chen Y, Mu D, Zhang H, Li N
\textit{et al}: \textbf{pIRS: Profile-based Illumina pair-end reads simulator}.
\textit{Bioinformatics }2012, \textbf{28}(11):1533-1535.}

{\selectlanguage{english}
24.~~~~~~ Lysholm F, Andersson B, Persson B: \textbf{An efficient simulator of
454 data using configurable }\textbf{statistical models}. \textit{BMC research
notes }2011, \textbf{4}(1):449.}

{\selectlanguage{english}
25.~~~~~~ Huang W, Li L, Myers JR, Marth GT: \textbf{ART: a next-generation
sequencing read simulator}. \textit{Bioinformatics }2012,
\textbf{28}(4):593-594.}

{\selectlanguage{english}
26.~~~~~~ McElroy KE, Luciani F, Thomas T: \textbf{GemSIM: general, error-model
based simulator of next-generation sequencing data}. \textit{BMC genomics
}2012, \textbf{13}:74.}

{\selectlanguage{english}
27.~~~~~~ Holtgrewe M: \textbf{Mason -- a read simulator for second generation
sequencing data.} Berlin: Freie Universität Berlin; 2010.}

{\selectlanguage{english}
28.~~~~~~ Richter DC, Ott F, Auch AF, Schmid R, Huson DH: \textbf{MetaSim: a
sequencing simulator for genomics and metagenomics}. \textit{PloS one }2008,
\textbf{3}(10):e3373.}

{\selectlanguage{english}
29.~~~~~~ Engle ML, Burks C: \textbf{Artificially generated data sets for
testing DNA sequence assembly algorithms}. \textit{Genomics }1993,
\textbf{16}(1):286-288.}

{\selectlanguage{english}
30.~~~~~~ Li H, Handsaker B, Wysoker A, Fennell T, Ruan J, Homer N, Marth G,
Abecasis G, Durbin R: \textbf{The Sequence Alignment/Map format and SAMtools}.
\textit{Bioinformatics }2009, \textbf{25}(16):2078-2079.}

{\selectlanguage{english}
31.~~~~~~ Amigo J, Salas A, Phillips C: \textbf{ENGINES: exploring single
nucleotide variation in entire human genomes}. \textit{BMC bioinformatics
}2011, \textbf{12}:105.}

{\selectlanguage{english}
32.~~~~~~ DePristo MA, Banks E, Poplin R, Garimella KV, Maguire JR, Hartl C,
Philippakis AA, del Angel G, Rivas MA, Hanna M \textit{et al}: \textbf{A
framework for variation discovery and genotyping using next-generation DNA
sequencing data}. \textit{Nature genetics }2011, \textbf{43}(5):491-498.}

{\selectlanguage{english}
33.~~~~~~ Mills RE, Pittard WS, Mullaney JM, Farooq U, Creasy TH, Mahurkar AA,
Kemeza DM, Strassler DS, Ponting CP, Webber C \textit{et al}: \textbf{Natural
genetic variation caused by small insertions and deletions in the human
genome}. \textit{Genome research }2011, \textbf{21}(6):830-839.}

{\selectlanguage{english}
34.~~~~~~ Amigo J, Phillips C, Salas A, Carracedo A: \textbf{Viability of
in-house datamarting approaches for population genetics analysis of SNP
genotypes}. \textit{BMC bioinformatics }2009, \textbf{10 Suppl 3}:S5.}

{\selectlanguage{english}
35.~~~~~~ Mullaney JM, Mills RE, Pittard WS, Devine SE: \textbf{Small insertions
and deletions (INDELs) in human genomes}. \textit{Human molecular genetics
}2010, \textbf{19}(R2):R131-136.}

{\selectlanguage{english}
36.~~~~~~ Redon R, Ishikawa S, Fitch KR, Feuk L, Perry GH, Andrews TD, Fiegler
H, Shapero MH, Carson AR, Chen W \textit{et al}: \textbf{Global variation in
copy number in the human genome}. \textit{Nature }2006,
\textbf{444}(7118):444-454.}

{\selectlanguage{english}
37.~~~~~~ Conrad DF, Pinto D, Redon R, Feuk L, Gokcumen O, Zhang Y, Aerts J,
Andrews TD, Barnes C, Campbell P \textit{et al}: \textbf{Origins and functional
impact of copy number variation in the human genome}. \textit{Nature }2010,
\textbf{464}(7289):704-712.}

{\selectlanguage{english}
38.~~~~~~ Mills RE, Walter K, Stewart C, Handsaker RE, Chen K, Alkan C, Abyzov
A, Yoon SC, Ye K, Cheetham RK \textit{et al}: \textbf{Mapping copy number
variation by population-scale genome sequencing}. \textit{Nature }2011,
\textbf{470}(7332):59-65.}

{\selectlanguage{english}
39.~~~~~~ Park H, Kim JI, Ju YS, Gokcumen O, Mills RE, Kim S, Lee S, Suh D, Hong
D, Kang HP \textit{et al}: \textbf{Discovery of common Asian copy number
variants using integrated high-resolution array CGH and massively parallel DNA
sequencing}. \textit{Nature genetics }2010, \textbf{42}(5):400-405.}

{\selectlanguage{english}
40.~~~~~~ Carter SL, Cibulskis K, Helman E, McKenna A, Shen H, Zack T, Laird PW,
Onofrio RC, Winckler W, Weir BA \textit{et al}: \textbf{Absolute quantification
of somatic DNA alterations in human cancer}. \textit{Nature biotechnology
}2012, \textbf{30}(5):413-421.}

{\selectlanguage{english}
41.~~~~~~ Ishikawa S, Komura D, Tsuji S, Nishimura K, Yamamoto S, Panda B, Huang
J, Fukayama M, Jones KW, Aburatani H: \textbf{Allelic dosage analysis with
genotyping microarrays}. \textit{Biochemical and biophysical research
communications }2005, \textbf{333}(4):1309-1314.}
\clearpage
\begin{figure}[h!]
\begin{flushleft}
\includegraphics[scale=0.7]{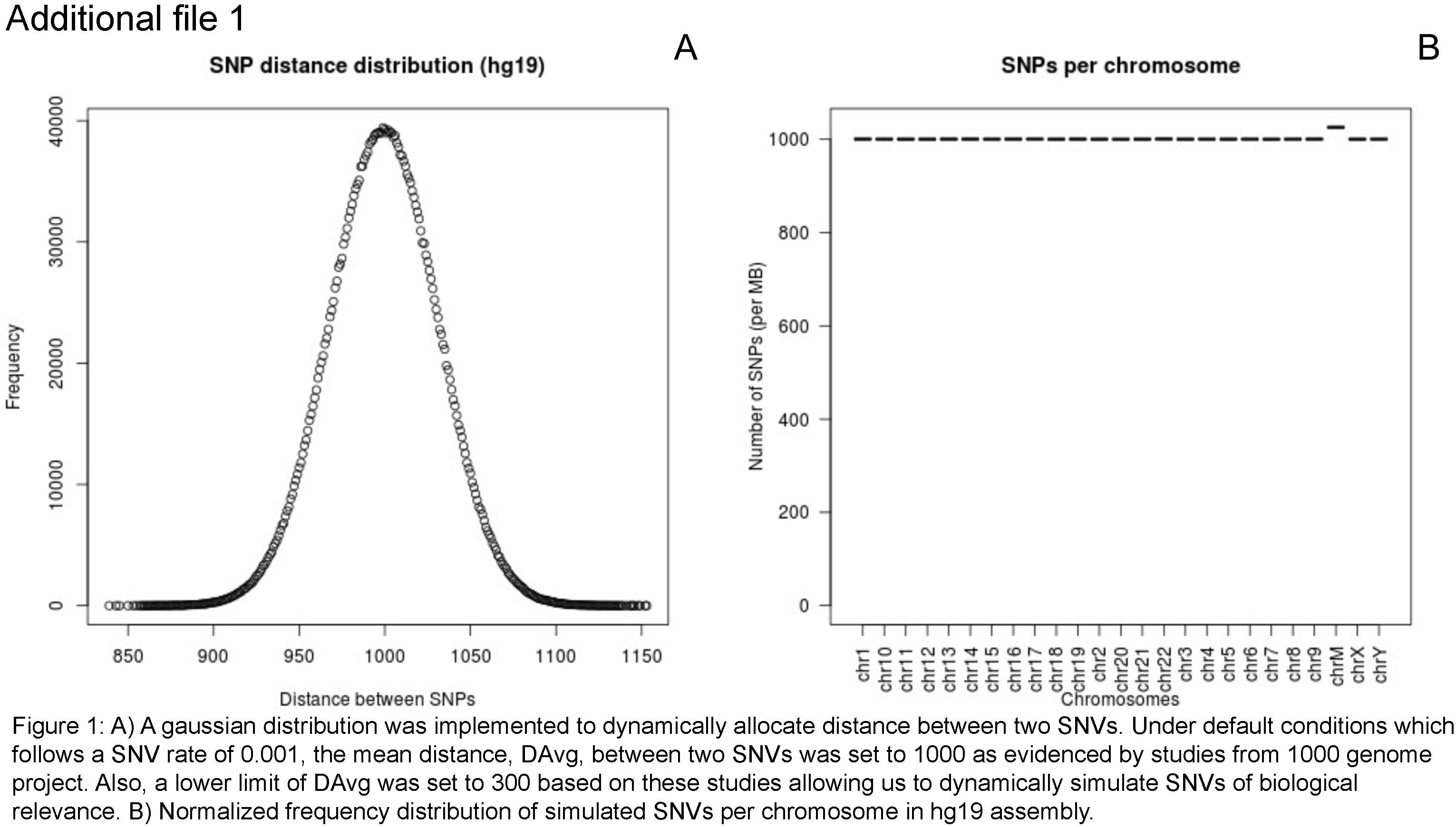}
\end{flushleft}
\end{figure}
\clearpage{\selectlanguage{english}\bfseries
Additional file 2}
\bigskip
\\
{\selectlanguage{english}
Time profile}
\begin{flushleft}
\tablehead{}
\begin{supertabular}{|m{0.44345984in}|m{0.96365983in}|m{0.54205986in}|m{0.38375986in}|m{0.40045986in}|}\hline
\selectlanguage{english} Tool & Version & \# cores & chr22 & hg19\\ \hline
\selectlanguage{english} pirs & 1.00 &
\raggedleft \selectlanguage{english} 1 &
\raggedleft \selectlanguage{english} 225 &
\raggedleft\arraybslash \selectlanguage{english} 5775\\ \hline
\selectlanguage{english} dwgsim &
\selectlanguage{english} 0.1.10 &
\raggedleft \selectlanguage{english} 1 &
\raggedleft \selectlanguage{english} 69 &
\raggedleft\arraybslash \selectlanguage{english} 15993\\ \hline
\selectlanguage{english} Gemsim &
\selectlanguage{english} 1.6 &
\raggedleft \selectlanguage{english} 1 &
\raggedleft \selectlanguage{english} 26778 &
\raggedleft\arraybslash \selectlanguage{english} 401640\\ \hline
\selectlanguage{english} art &
\selectlanguage{english} banana\_packages &
\raggedleft \selectlanguage{english} 1 &
\raggedleft \selectlanguage{english} 130 &
\raggedleft\arraybslash \selectlanguage{english} 11718\\ \hline
\selectlanguage{english} SInC &
\selectlanguage{english} 1.00 &
\raggedleft \selectlanguage{english} 1 &
\raggedleft \selectlanguage{english} 123 &
\raggedleft\arraybslash \selectlanguage{english} 3762\\ \hline
\selectlanguage{english} SInC &
\selectlanguage{english} 1.00 &
\raggedleft \selectlanguage{english} 2 &
\raggedleft \selectlanguage{english} 86 &
\raggedleft\arraybslash \selectlanguage{english} 3124\\ \hline
\selectlanguage{english} SInC &
\selectlanguage{english} 1.00 &
\raggedleft \selectlanguage{english} 3 &
\raggedleft \selectlanguage{english} 84 &
\raggedleft\arraybslash \selectlanguage{english} 3150\\ \hline
\selectlanguage{english} SInC &
\selectlanguage{english} 1.00 &
\raggedleft \selectlanguage{english} 4 &
\raggedleft \selectlanguage{english} 94 &
\raggedleft\arraybslash \selectlanguage{english} 3285\\ \hline
\end{supertabular}
\end{flushleft}
{\selectlanguage{english}
SNP re-discovery}
\begin{flushleft}
\tablehead{}
\begin{supertabular}{|m{0.42405984in}|m{0.49625984in}|m{0.6504598in}|m{0.63445984in}|m{0.72055985in}|m{0.7483598in}|}\hline
\selectlanguage{english} Tool &
\selectlanguage{english} coverage &
\selectlanguage{english} \# SNPs simulated &
\selectlanguage{english} \%redisc PASS &
\selectlanguage{english} \%het PASS redisc &
\selectlanguage{english} \%hom PASS redisc\\ \hline
\selectlanguage{english} pirs &
\selectlanguage{english} 20.56 &
\selectlanguage{english} 35150 &
\selectlanguage{english} 97.19 &
\selectlanguage{english} NA &
\selectlanguage{english} NA\\ \hline
\selectlanguage{english} dwgsim &
\selectlanguage{english} 19.46 &
\selectlanguage{english} 31291 &
\selectlanguage{english} 86.53 &
\selectlanguage{english} 80.72 &
\selectlanguage{english} 97.44\\ \hline
\selectlanguage{english} gemsim &
\selectlanguage{english} 21.75 &
\selectlanguage{english} 23822 &
\selectlanguage{english} 84.68 &
\selectlanguage{english} NA &
\selectlanguage{english} NA\\ \hline
\selectlanguage{english} SInC &
\selectlanguage{english} 19.48 &
\selectlanguage{english} 34914 &
\selectlanguage{english} 98.68 &
\selectlanguage{english} 97.29 &
\selectlanguage{english} 99.00\\ \hline
\end{supertabular}
\end{flushleft}
{\selectlanguage{english}
Indel re-discovery}
\begin{flushleft}
\tablehead{}
\begin{supertabular}{|m{0.41565984in}|m{0.49625984in}|m{0.50185986in}|m{0.7504598in}|m{0.7066598in}|}
\hline
\selectlanguage{english} Tool &
\selectlanguage{english} coverage &
\selectlanguage{english} INDEL size &
\selectlanguage{english} \# INDELs simulated &
\selectlanguage{english} \% INDELs redisc\\ \hline
\selectlanguage{english} pirs &
\selectlanguage{english} 22.03 &
\selectlanguage{english} all &
\selectlanguage{english} 42877 &
\selectlanguage{english} 75.64\\ \cline{3-5}
 &
 &
\selectlanguage{english} 1-6 &
\selectlanguage{english} 42848 &
\selectlanguage{english} 75.77\\ \cline{3-5}
 &
 &
\selectlanguage{english} 7-10 &
\selectlanguage{english} 0 &
\selectlanguage{english} 0.00\\ \cline{3-5}
 &
 &
\selectlanguage{english} 11-20 &
\selectlanguage{english} 0 &
\selectlanguage{english} 0.00\\ \cline{3-5}
 &
 &
\selectlanguage{english} 20- &
\selectlanguage{english} 29 &
\selectlanguage{english} 37.93\\ \hline
\selectlanguage{english} dwgsim &
\selectlanguage{english} 19.46 &
\selectlanguage{english} all &
\selectlanguage{english} 42388 &
\selectlanguage{english} 44.03\\ \cline{3-5}
 &
 &
\selectlanguage{english} 1-6 &
\selectlanguage{english} 42375 &
\selectlanguage{english} 57.04\\ \cline{3-5}
 &
 &
\selectlanguage{english} 7-10 &
\selectlanguage{english} 13 &
\selectlanguage{english} 23.08\\ \cline{3-5}
 &
 &
\selectlanguage{english} 11-20 &
\selectlanguage{english} 0 &
\selectlanguage{english} 0.00\\ \cline{3-5}
 &
 &
\selectlanguage{english} 20- &
\selectlanguage{english} 0 &
\selectlanguage{english} 0.00\\ \hline
\selectlanguage{english} SInC &
\selectlanguage{english} 19.87 &
\selectlanguage{english} all &
\selectlanguage{english} 50403 &
\selectlanguage{english} 91.83\\ \cline{3-5}
 &
 &
\selectlanguage{english} 1-6 &
\selectlanguage{english} 44659 &
\selectlanguage{english} 92.08\\ \cline{3-5}
 &
 &
\selectlanguage{english} 7-10 &
\selectlanguage{english} 4273 &
\selectlanguage{english} 90.73\\ \cline{3-5}
 &
 &
\selectlanguage{english} 11-20 &
\selectlanguage{english} 514 &
\selectlanguage{english} 90.27\\ \cline{3-5}
 &
 &
\selectlanguage{english} 20- &
\selectlanguage{english} 957 &
\selectlanguage{english} 65.10\\ \cline{3-5}
 &
 &
\selectlanguage{english} 21-50 &
\selectlanguage{english} 386 &
\selectlanguage{english} 82.12\\ \hline
\end{supertabular}
\end{flushleft}
\clearpage
\begin{figure}[h!]
\begin{flushleft}
\includegraphics[scale=0.7]{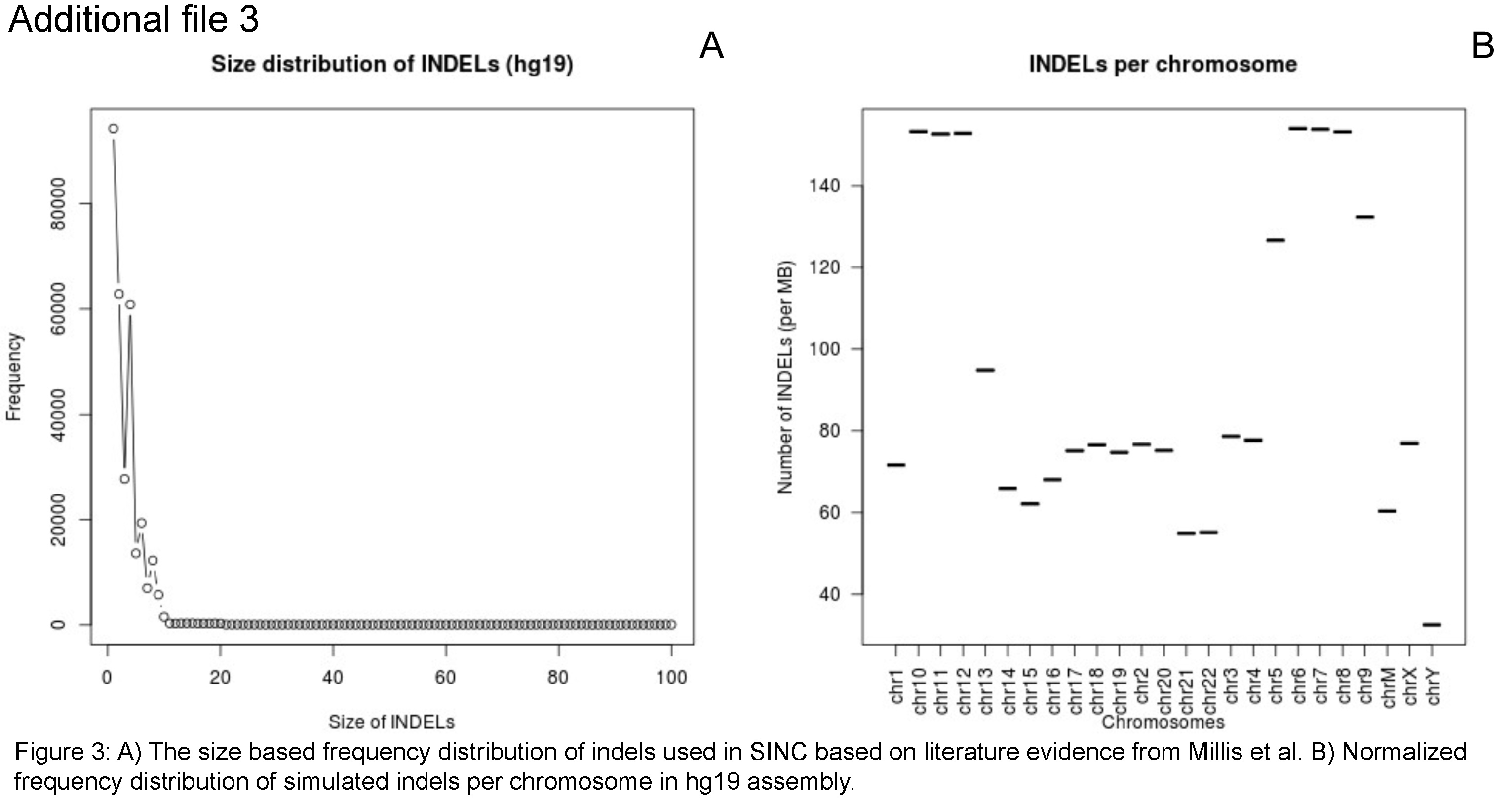}
\end{flushleft}
\end{figure}
\clearpage
\begin{figure}[h!]
\begin{flushleft}
\includegraphics[scale=0.5]{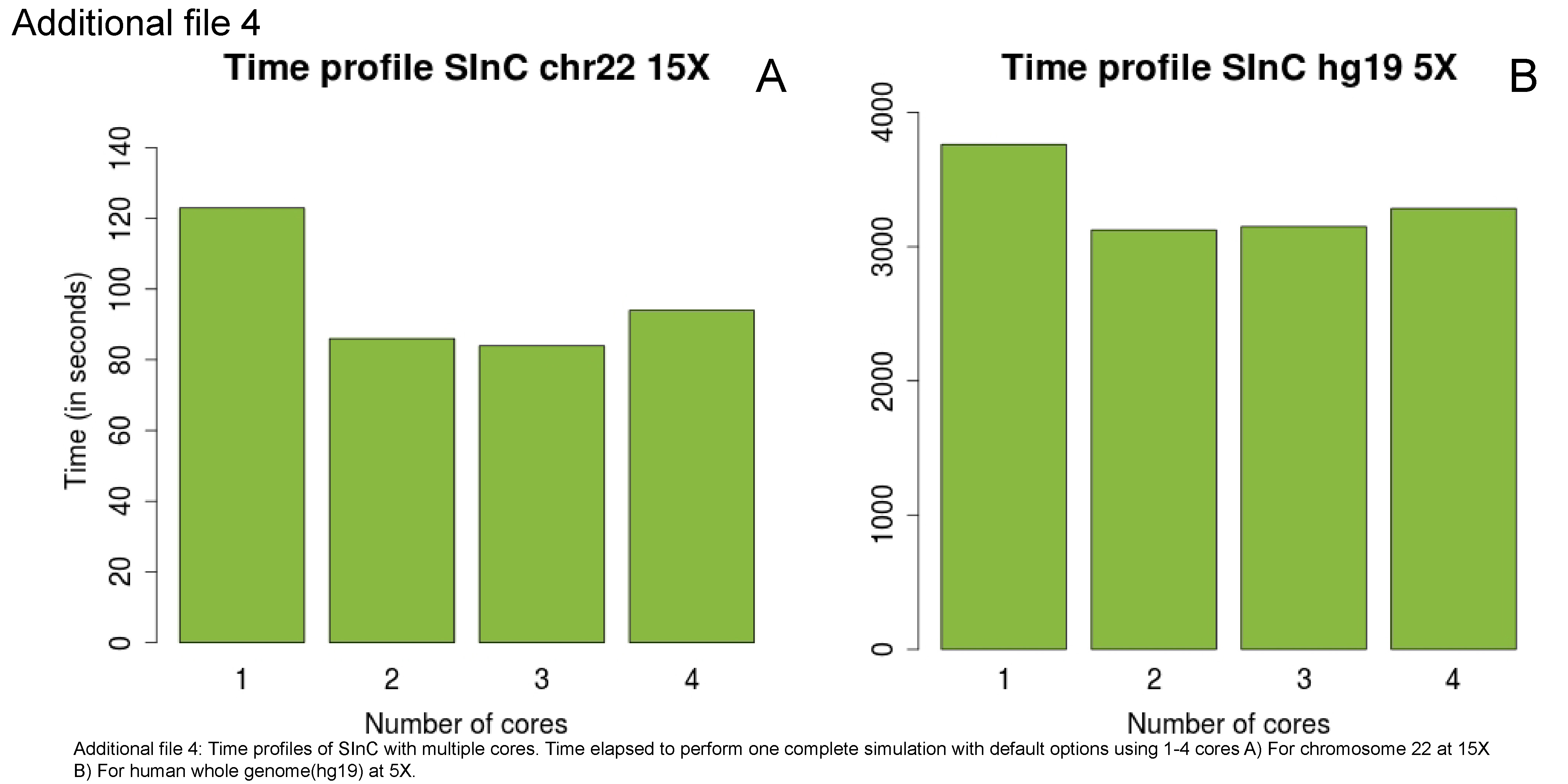}
\end{flushleft}
\end{figure}
\end{document}